# Shell-resolved melting kinetics of an icosahedral cluster

*Hong H. Liu[1], En Y. Jiang[1,*], Hai L. Bai[1], Ping Wu[1], Zhiqing Li[1] and Chang Q. Sun[2]*

[1]Tianjin Key Laboratory of Low Dimensional Materials Physics and Preparing Technology,

Institute of Advanced Materials Physics, Faculty of Science, Tianjin University, Tianjin 300072,

People's Republic of China

[2]School of Electrical and Electronic Engineering, Nanyang Technological University, Singapore

639798

Molecular dynamics calculations of the fluctuation of bond vibration revealed the shell-resolved kinetics of surface melting of closed-shelled cluster containing 147atoms with Lennard-Jones type interaction. It is found that the surface melting is imitated by the migrating of the vertex atoms and the melting process can be divided into three major stages, i.e., vertex migrating, surface melting, and general melting. Although the melting process of the LJ147 cluster could be divided into discrete stages of shell-by-shell surface melting, in general, there is still a continuous process of melting from the surface to the core interior.

- Corresponding author. E-mail: eyjiang@tju.edu.cn; ecqsun@ntu.edu.sg

## 1. Introduction

The thermal stability of nanostructures [1,2,3,4,5,6,7,8,9] and surface melting [10,11,12,13,14,15] are of great importance to the fundamental knowledge and to the upcoming technological applications, which has attracted tremendous interest recently. The lowered thermal stability may hinder the advancement of nanoscale devices and hence under standing the kinetics is necessary. Although the melting mechanism of clusters of nanometers range or larger has been well understood in terms of the liquid drop model [16], lattice and surface phonon instability model [17], surface atom vibration model [18,19,20], surface area difference related cohesive energy model [21], and the bond-order-length-strength (BOLS) model [22,23], the dynamics and kinetics of melting of clusters containing a few accountable atoms is yet to be understood. The adatoms or atomic cavities on the cluster surface play an important role in the surface melting [24,25] because of the lowered atomic cohesive energy [22]. Using the atom-resolved computations, Lee *et al* [26] suggested that the potential energy distribution of the atoms has strong effect on the melting of clusters. In the simulations of 561- and 923-atom Mackay icosahedra, Doye *et al* [27] derived that at temperature below the melting point, vacancies are generated at the vertices and edges through atomic diffusion from the interior to the surfaces, because of surface reconstruction rather than surface melting.

The atom-resolved simulation could provide detailed information about the change of structure and energy of the small clusters. However, for a cluster containing hundreds of atoms, it is difficult to distinguish and trace the individual atom at the surface or in the cluster interior. Therefore, previous simulating studies of surface melting were focused on clusters containing several or tens of atoms [24,25,26]. Cheng *et al* [28] divided the atoms in the clusters with 40-147 atoms into subregions according to the distances between the mass center of the cluster and the atoms of concern. However, in the previous work [29], we have shown that an alternative scheme is more proper for distinguishing the surface shell of a clubbed cluster (such as 19-atom Lennard-Jones double-icosahedron cluster). Basing on a tangential plane method, we divided the clusters into atomic shells. We found that the outermost shells are so important that they can change the structure and potential energy of the entire cluster through the surface strain induced compressive stress. In the current work, we report the study of the melting kinetics of a 147-atom Lennard-Jones cluster (LJ147) through a shell-resolved Molecular Dynamics (MD) simulation. Results revealed that for the icosahedral LJ147 cluster, the surface melting is initiated by the migration of the vertex atoms.

**2. Principles and method**

Verlet algorithm [30] was adopted in the MD simulation. The atoms are bound by the Lennard-Jones pair potential

$$P(r_{ij}) = 4\varepsilon\left[\left(\sigma / r_{ij}\right)^{12} - \left(\sigma / r_{ij}\right)^{6}\right] \quad , \tag{1}$$

where $r_{ij}$ is the distance between atom $i$ and $j$. $\varepsilon$ and $\sigma$ are used as the unit for energy and length, respectively. We may divide the LJ147 cluster into three shells [29] (shell-$s$, $s$ = 0 ~ 3 counts from the center atom to the outermost and shell-0 denotes the center atom of the icosahedron). In the calculation, the shell-resolved quantities such as atomic mean kinetic energy $\langle ek_s \rangle_t$ and atomic mean potential energy $\langle ep_s \rangle_t$ of atoms in the given shell were calculated,

$$\langle ek_s \rangle_t = \left\langle \frac{1}{N_s} \sum_{i \in shell-s} ek_{atom-i} \right\rangle_t$$

$$\langle ep_s \rangle_t = \left\langle \frac{1}{N_s} \sum_{i \in shell-s} ep_{atom-i} \right\rangle_t ,$$

the subscript $t$ denotes the time average, $N_s$ is the atom number of shell-*s*. $ek_{atom-i}$ and $ep_{atom-i}$ are the kinetic and potential energy of the ith atom in shell-*s*, respectively. The shell-resolved root-mean-square bond length fluctuation $\langle \delta_s \rangle_t$ [5] is calculated for monitoring the phase changes of different shells:

$$\langle \delta_s \rangle_t = \frac{2}{N_s(N_s-1)} \sum_{i,j \in shell-s, i>j} \frac{\left( \langle r_{ij}^2 \rangle_t - \langle r_{ij} \rangle_t^2 \right)^{1/2}}{\langle r_{ij} \rangle_t},$$

$r_{ij}$ is the distance between atom $i$ and $j$ in shell-*s*. The structure of the LJ147 cluster was built up using the method of Xiang *et al* [31]. The internal temperature being related to the kinetic energy of the cluster is given by

$$T = \frac{2\langle E_k \rangle_t}{k_B(3N-6)} ,$$

where $\langle E_k \rangle_t$ is the time average of the total kinetic energy of the cluster along the whole trajectory, and $k_B$ is the Boltzmann constant, then $\varepsilon / k_B$ is the unit of temperature. A velocity scaling method was adopted as the temperature is raised by a step of $\Delta T = 0.01\varepsilon / k_B$. At each point of temperature, an adiabatic process of $3\times10^6$ MD steps was introduced in which $\langle ek_s \rangle_t$, $\langle ep_s \rangle_t$, and $\langle \delta_s \rangle_t$ were calculated.

**3. Results and discussion**

Results shown in Fig. 1a indicates that the mean atomic kinetic energies $\langle ek_s \rangle_t$ in different shells are statistically identical when the LJ147 cluster is in a solid-like state. This finding suggests that for a correlated many-body system in thermal equilibrium, the mean kinetic energy of all the atoms is independent of the surroundings. It is known that the potential energy of an atom depends on the position of this atom [Eq. 1], so the mean atomic potential energy $\langle ep_s \rangle_t$ in each shell shown in Fig. 1b is quite different. Because the atoms at the surface have less coordinates than the inner ones, so they are weakly bound, and then they have higher potential energy (Fig. 1b and Eq. 1) than the inner ones in their thermal motion. Therefore, when the temperature is raised, the atoms on the surface will readily gain thermal energy for vibration and structure transition, as can be seen from Fig. 1c. As the temperature is raised, the $\langle \delta_s \rangle_t$ of the surface shell reaches the Lindermann's criteria first ($\langle \delta_s \rangle_t > 0.10$) [5]. The splitting of the $\langle \delta_s \rangle_t$ curves of different shells in Fig. 1c shows clearly that the surface shell melts first at a temperature that is slightly lower than that for the inner shells. It should be noted here that when the whole cluster starts melting, the distinction between different shells in Fig. 1 is no longer essential.

It could be noted that, for the LJ147 cluster, although the shell-0 atom has the same coordination number (CN) (CN=12) with the atoms in shell-1 and shell-2, the potential energy of the innermost ones is higher than that of atoms in shell-1 and shell-2 [Fig. 1b], due to the compression from the outer shells [29]. It could be seen from Fig. 1b that, when the cluster is in the solid-like state at elevated temperature, the potential energy of the shell-0 atom becomes even lower rather than higher as one expect, because of the structural relaxation and the weakening of the compression on the shell-0. A similar phenomenon has also been found by Lee *et al* [26], in which the potential energy of the internal atoms in the $Pd_{19}$ of double icosahedron is lowered by ejecting one of the inner atoms to the surface of the cluster when the temperature is raised.

In the calculation, we have found that the surface melting of the icosahedral LJ147 cluster is initiated by the migration of atoms at the vertexes of the surface [Fig. 2]. The atoms leave the vertexes and become capping atoms, floating over the surface of the cluster [Figs. 2b-c]. As the energy of the cluster becomes higher, more vertex atoms leave their original positions and become capping ones [Figs. 2d-h]. When $T_i = 0.205$, the surface shell generally melts, the capping atoms are no longer distinguishable, so no capping atom is denoted for Fig. 2i.

This phenomenon generally extends the understanding on surface melting [25], which suggests that the surface melting is aroused from the migration of the capping atoms or the vacancies on the surface. For an icosahedral cluster without adherent capping atoms, the surface melting could also be aroused from the capping atoms translated from migrating vertex ones. It could be noticed that the CN of the vertex atoms is 6, smaller than that of other surface atoms (8 for the atoms on the edges and 9 for those on the faces, respectively). We should note here that because most of the surface atoms reside in the low-coordinate edge and vertex sites of the LJ147 cluster [27], the disordering aroused from the migration of the under-coordinated atoms will affect the whole structural integrity of the surface shell, and the reconstruction of the surface shell could not be observed. This phenomenon could be understood in terms of the BOLS correlation mechanism [23,32], which indicates that the bond order loss of an atom causes the remaining bonds of the under-coordinated atom to contract spontaneously associated with bond strength gain, resulting in a depression of atomic cohesive energy, generally. Therefore, the melting temperature varies from site to site depending on the product of the effective atomic CN and the cohesive energy per bond of the less-coordinated atom. So the under-coordinated vertex atoms could be thermally activated more readily than the ones in the surface plane. The migration of the vertex atoms will produce more vacancies around which atoms loss their coordinates. Therefore, the migration of the vertex atoms initiates the surface melting at temperature lower than that for global melting. Lee *et al* [26] explained reasonably the surface melting of small clusters using the parameter of $S = \overline{E}_{\text{int}} / \overline{E}_{sur} \geq 1$ with $\overline{E}_{\text{int}}$ and $\overline{E}_{sur}$ being the mean potential energies of the internal and surrounding atoms, respectively. The BOLS correlation in terms of atomic cohesive energy may represent well the situation discussed in comparison to using the parameter $S$ in explaining the surface melting aroused from the migration of the vertex atoms over the surface. The former uses the concept of cohesive energy of the specific atoms and the latter uses the concept of average potential energy.

Fig. 3 shows that when $T_d = 0.205$, most of the surface (shell-3) atoms start migrating and exchanging their positions, which denotes the second stage of cluster melting, or the general melting of the surface, while the inner shells still keep their solid-like state. However, we noticed that a few atoms jumped from shell-2 to the shell 3. At the temperature of surface melting, some inner atoms, such as the ones surrounding the vertex vacancies, also join the surface shell due to fluctuation. This phenomenon implies that for a cluster willing to show surface melting, the melting proceeds from surface to the inner in a continuous way rather than concisely the shell-by-shell wise. When $T_f = 0.211$, the two inner shells have lost their regular configurations and more atoms in shell-2 exchange their positions with the surface atoms, but no shell-1 atoms are found on the surface. Until $T_g = 0.225$, the cluster is generally melted and the atoms of all the three shells could be found on the surface. This global melting could be observed in Fig. 1c as the $\langle \delta_s \rangle_t$ curves for the three shells become no longer distinguishable. Although the cluster is generally melted at $T_g = 0.225$, some surface atoms have evaporated at $T_d = 0.205$ when surface melting is dominant.

## 4. Conclusion

We have found that the surface melting of small cluster could be initiated from the migration of the vertex atom, in which the potential energy of the atoms plays more important role than the kinetic energy. The compression aroused from the outer atomic shells is found to affect the thermal dynamics of the entire cluster. The shell-resolved $\langle \delta_s \rangle_t$ could be considered to be an effective variable for detecting the surface melting. Although the melting of LJ147 cluster could be divided into three major stages (vertex migration, surface melting, and general melting), the process of melting and evaporating proceeds in a continuous way rather than concisely the shell-by-shell wise.

Figure captions

Fig. 1 Temperature dependence of the shell-resolved properties of the LJ147 atomic cluster. (a) Atomic mean kinetic energy; (b) atomic mean potential energy; and (c) the shell-resolved root-mean-square bond length fluctuation.

Fig. 2 Atomic migration kinetics on the surface of the LJ147 cluster at various temperatures. The vertex atoms (in black) are more easily to leave their original positions and move over the surface of the cluster. Panels (a-i) denote the time order in simulation. The temperatures of each panel from $T_a$ to $T_i$ are: 0.165, 0.169, 0.169, 0.190, 0.190, 0.190, 0.190, 0.203, 0.205 ($\varepsilon / k_B$), respectively. When $T_i = 0.205$, some atoms are evaporated from the surface shell of the cluster, and this shell has generally melted, so there is no atom in black for panel-i.

Fig. 3 The melting process of the LJ147 cluster exhibited by configurations of the three shells at different temperatures. The atoms of different shells are in different colors. Rows (a-h) denote the time order in simulation. The temperatures of each row from $T_a$ to $T_h$ are 0.156, 0.182, 0.190, 0.205, 0.209, 0.211, 0.225, 0.241 ($\varepsilon / k_B$), respectively. Clusters in the left column are the configurations of the whole cluster.

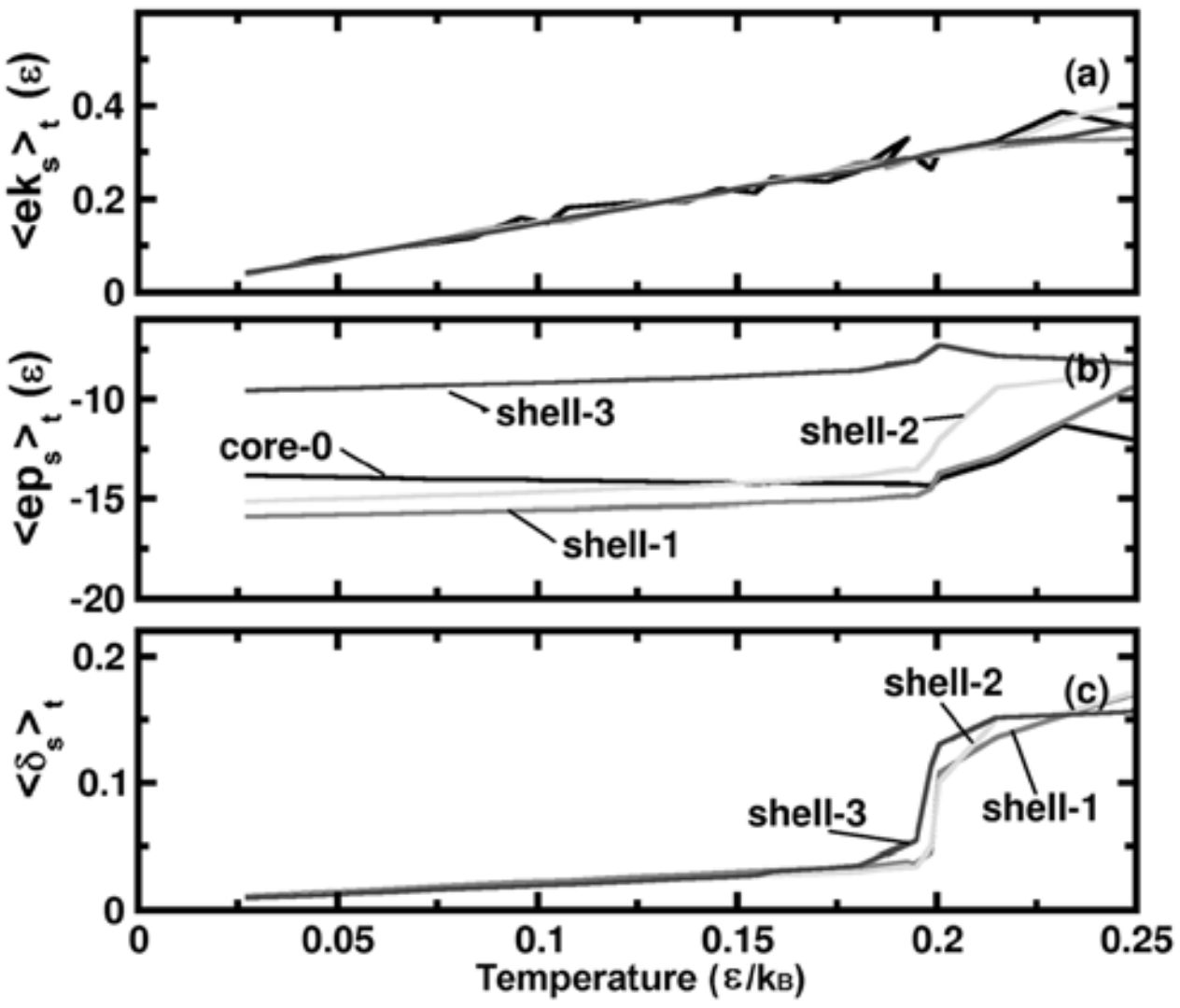

Fig 1

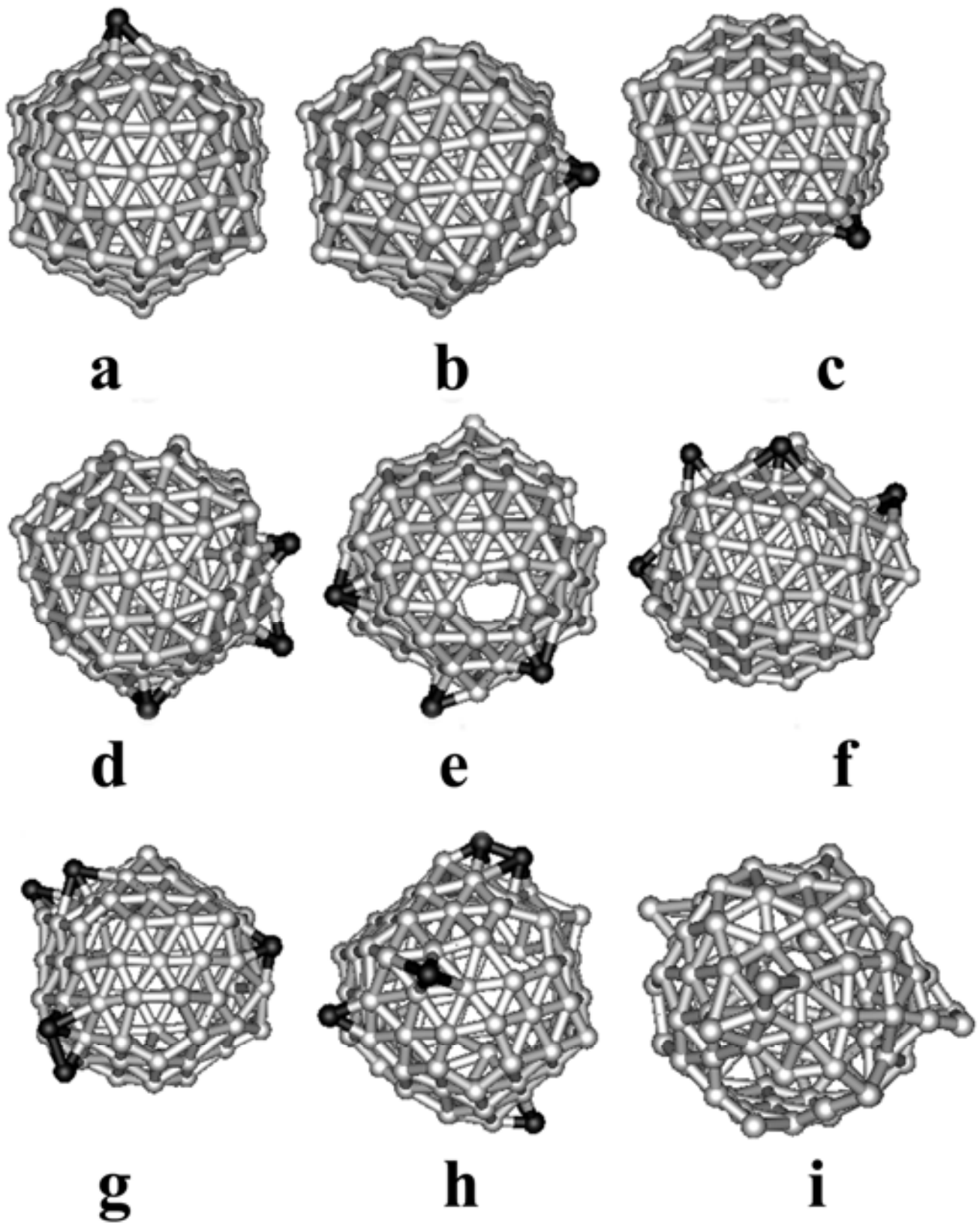

Fig 2

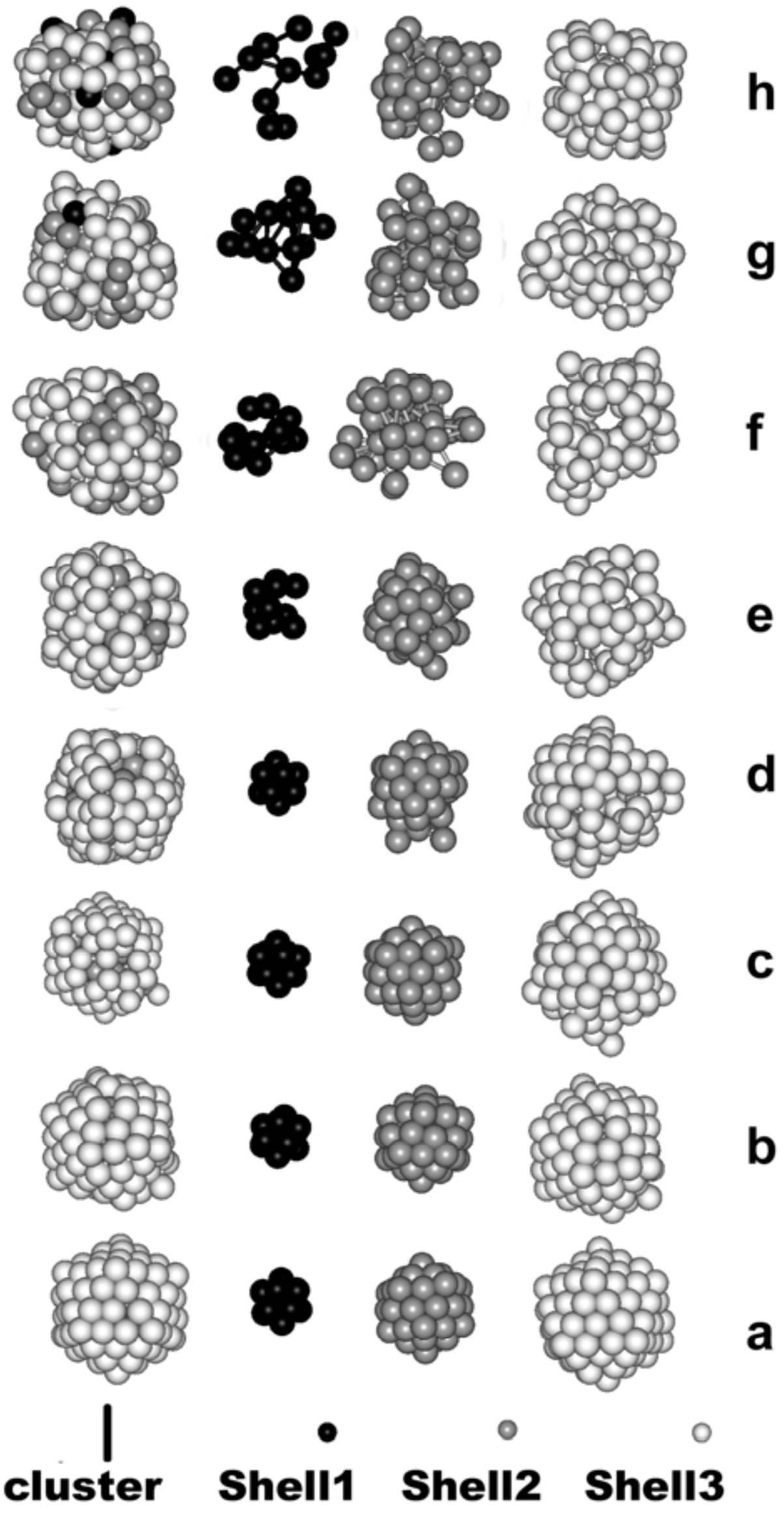

Fig 3

**Reference**

[1] Schmidt, M.; Kusche, R.; von Issendorff, B.; Haberland, H. *Nature* **1998**, 393, 238.

[2] Chacko, S.; Kanhere, D. G.; Blundell, S. A. *Phys. Rev. B* **2005**, 71, 155407.

[3] Lai, S. L.; Guo, J. Y.; Petrova, V.; Ramanath, G.; Allen, L. H. *Phys. Rev. Lett.* **1996**, 77, 99.

[4] Cleveland, C. L.; Luedtke W. D.; Landman, U. *Phys. Rev. Lett.* **1998**, 81, 2036.

[5] Arslan, H.; Güven, M. H. *New J. Phys.* **2005**, 7, 60.

[6] Shvartsburg, A. A.; Jarrold, M. F. *Phys. Rev. Lett.* 85 (2000) 2530.

[7] Haberland, H.; Hippler, T.; Donges, J.; Kostko, O.; Schmidt, M.; von Issendorff, B. *Phys. Rev. Lett.* **2005**, 94, 035701.

[8] Hendy, S. C. *Phys. Rev. B* **2005**, 71, 115404.

[9] Ding, F.; Rosén, A.; Bolton, K. *Phys. Rev. B* **2004**, 70, 075416.

[10] Breaux, G. A.; Neal, C. M.; Cao, B.; Jarrold, M. F. *Phys. Rev. Lett.* **2005**, 94, 173401.

[11] Lai, S. K.; Lin, W. D.; Wu, K. L.; Li, W. H.; Lee, K. C. *J. Chem. Phys.* **2004**, 121, 1487.

[12] Calvo, F.; Spiegelman, F. *J. Chem. Phys.* **2004**, 120, 9684.

[13] Joshi, K.; Kanhere, D. G.; Blundell, S. A. *Phys. Rev. B* **2003**, 67, 235413.

[14] Calvo, F.; Spiegelmann, F. *J. Chem. Phys.* **2000**, 112, 2888.

[15] Aguado, A.; López, J. M.; Alonso, J. A.; Stott, M. J.; *J. Chem. Phys.* **1999**, 111, 6026.

[16] Nanda, K.K.; Sahu, S.N. and Behera, S.N. Phys Rev A 2002;66:013208.

[17] Wautelet M, Phys Lett 1998, A 246, 341.

[18] Jiang, Q.; Liang, L. H.; Zhao D. S. J Phys Chem B 2001,105, 6275.

[19] Jiang, Q.; Shi, H.X.; Zhao, M. J Chem Phys 1999,111,2176.

[20] Shi, F. G. J. Mater Res 1994, 9,1307.

[21] Qi, W.H.; Wang, M.P.; Xu, G.Y. Chem Phys Lett. 2003, 372, 632.

[22] Sun, C.Q.; *Prog Solid State Chem* 2007, 35, 1.

[23] Sun, C. Q.; Wang, Y.; Tay, B. K.; Li, S.; Huang, H.; Zhang, Y. B. *J. Phys. Chem. B*

**2002**, 106, 10701.

[24] Sebetci, A.; Guvenc, Z. B. *Modelling Simul. Mater. Sci. Eng.* **2004**, 12, 1131.

[25] Lee, Y. J.; Maeng, J. Y.; Lee, E.-K.; Kim, B.; Kim, S.; Han., K.-K. *J. Comput. Chem.* **2000**, 21, 380.

[26] Lee, Y. J.; Lee, E.-K.; Kim S.; Nieminen, R. M. *Phys. Rev. Lett.* **2001**, 86, 999.

[27] Doye, J. P. K.; Wales, D. *J. Z. Phys. D* **1997**, 40, 466.

[28] Cheng, H.-P.; Berry, R. S. *Phys. Rev. A* **1992**, 45, 7969.

[29] Liu, H. H.; Jiang, E. Y.; Bai, H. L.; Wu, P.; Li, Z. Q. *Chem. Phys. Lett.* **2005**, 412, 195.

[30] Pang, T. An Introduction to Computational Physics, Cambridge University Press, Cambridge, UK, 1997.

[31] Xiang, Y.; Jiang, H.; Cai, W.; Shao, X. *J. Phys. Chem. A* **2004**, 108, 3586.

[32] Sun, C. Q.; Shi, Y.; Li, C. M.; Li, S.; Au Yeung, T. C. *Phys Rev. B* **2006**, **73**, 075408.